\documentclass[aps,prl,linenumbers,twocolumn]{revtex4-1}

\usepackage{epsfig}
\usepackage{graphicx}  
\usepackage{bm}        
\usepackage{amssymb}   
\usepackage[usenames, dvipsnames]{color}

\usepackage[format=plain,justification=centerlast,labelfont=bf]{caption}
\usepackage[caption=false]{subfig}

\renewcommand{\figurename}{Figure}

\begin{document}
\nolinenumbers

\title{
 Deterministic phase slips in mesoscopic superconducting rings
}

\author{ I. Petkovi\'{c}$^{1}$, A. Lollo$^{1}$, L.I. Glazman$^{1,2}$, and
J.G.E. Harris$^{1,2}$
}

\affiliation{
 $^{1}$  Department of Physics, Yale University, 217 Prospect St, New Haven, Connecticut 06520, USA\\
  $^{2}$ Department of Applied Physics, Yale University, 15 Prospect St, New Haven, Connecticut 06520, USA \\
    }

\begin{abstract}
\noindent

The properties of one-dimensional superconductors are strongly influenced by topological fluctuations of the order parameter, known as phase slips, which cause the decay of persistent current in superconducting rings and the appearance of resistance in superconducting wires. Despite extensive work, quantitative studies of phase slips have been limited by uncertainty regarding the order parameter's free energy landscape. Here we show detailed agreement between measurements of the persistent current in isolated flux-biased rings and Ginzburg-Landau theory over a wide range of temperature, magnetic field, and ring size; this agreement provides a quantitative picture of the free energy landscape. We also demonstrate that phase slips occur deterministically as the barrier separating two competing order parameter configurations vanishes. These results will enable studies of quantum and thermal phase slips in a well-characterized system and will provide access to outstanding questions regarding the nature of one-dimensional superconductivity.

\end{abstract}

\pacs{}
\maketitle

\pagestyle{empty}

\email{ ivana.petkovic@yale.edu}


Phase slips are topological fluctuations of the order parameter in one-dimensional superconductors \cite{little}.
They are responsible for the emergence of finite resistance in the superconducting state and for the decay of supercurrent in a closed loop \cite{langer_ambegaokar,mccumber_halperin,tinkham}.
Despite extensive research and a good understanding of their basic features, there remain a number of open questions related to their dynamics \cite{halperin}. One of the conceptually simplest systems in which to study phase slips is an isolated, flux-biased ring. Such a system can access several metastable states, and undergoes a phase slip when it passes from one of these states to another \cite{langer_ambegaokar}.
Tuning the free energy barrier between the states to zero with the applied flux $\Phi$ will result in a deterministic phase slip from the state that has become unstable \cite{tarlie_elder}, while tuning the barrier to a small but non-zero value will lead to a stochastic phase slip via thermal activation \cite{langer_ambegaokar,mccumber_halperin} or quantum tunneling \cite{giordano,duan,zaikin,golubov_zaikin,bradley_doniach,fazio_vanderzant,matveev,buchler}.

The interpretation of measurements of stochastic phase slips \cite{giordano,taps_exp,qps_exp1,qps_exp2,qps_exp3,qps_exp7,qps_exp8,qps_exp9,qps_exp10,artyunov,belkin_tech,belkin} has been complicated by these processes' strong dependence on the system's details, such as the form of the free energy landscape, the damping of the order parameter, and the noise driving its   fluctuations. Of particular importance is accurate knowledge of the barrier between metastable states, which enters exponentially into the rate of stochastic phase slips \cite{halperin}. In contrast, deterministic phase slips are governed solely by the form of the free energy landscape: they occur when the barrier is tuned to zero. For a strictly one-dimensional ring   (in which the order parameter only varies along the ring's circumference), Ginzburg-Landau (GL) theory can be used to analytically calculate the barrier height, the flux at which the deterministic phase slips occur \cite{langer_ambegaokar,mccumber_halperin},
and the measurable properties of the metastable states, e.g. their persistent current \cite{langer_ambegaokar,mccumber_halperin,zhang} and heat capacity \cite{bourgeois}.
 As a result, measurements of these properties that demonstrate precise agreement with
  theory are important for benchmarking a system in which to study thermal and quantum stochastic phase slips.
Previous measurements of persistent current $I(\Phi)$ in isolated superconducting rings have found quantitative agreement with
theory only at low magnetic field and very close to the critical temperature $T_\mathrm{c}$, where metastability is absent or nearly absent \cite{bert_moler,koshnick_moler,zhang}. However at lower temperatures, where metastability is well-established, only qualitative agreement with theory has been demonstrated \cite{pedersen,vodolazov, bluhm_moler}.

Here we present measurements of $I(\Phi)$ in isolated superconducting rings for temperatures spanning $T_\mathrm{c}/2<T<T_\mathrm{c}$. The results, over the full range of magnetic field, show quantitative agreement with the GL theory augmented by the empirical two-fluid model~\cite{tinkham}; the latter states the temperature dependence of the input parameters of GL theory in a broad temperature domain. The combination of the GL theory, nominally valid only at $T\to T_\mathrm{c}$, with the two-fluid model has been shown to accurately represent the results of microscopic theory down to $T\approx T_\mathrm{c}/2$ and was successfully used, e.g. in explaining measurement of the parallel critical field of thin Al films~\cite{tedrow_meservey} in this temperature range.
We find that phase slips occur at the flux values predicted by GL theory, even to the point of demonstrating a small correction due to the rings' finite circumference \cite{kramer_zimmermann,tuckerman}. In addition, we find that the dynamics of the phase slips is strongly damped, so that the disappearance of a barrier leads the system to relax to the adjacent local minimum. The measurement described here employs cantilever torque magnetometry, which has been shown to be a minimally-invasive probe of persistent current in isolated metal rings \cite{ania_science} and is capable of resolving individual phase slips in a single ring \cite{will_thesis}.
As a result these measurements demonstrate the essential features for studying stochastic phase slips: samples with a well-characterized free energy landscape, and a detection scheme suitable for measuring their intrinsic dynamics.

\vspace{3mm}
\noindent \textbf{\large{Results}}

\begin{figure}[h!]
\vspace{2mm} \centerline{\hbox{
\epsfig{figure=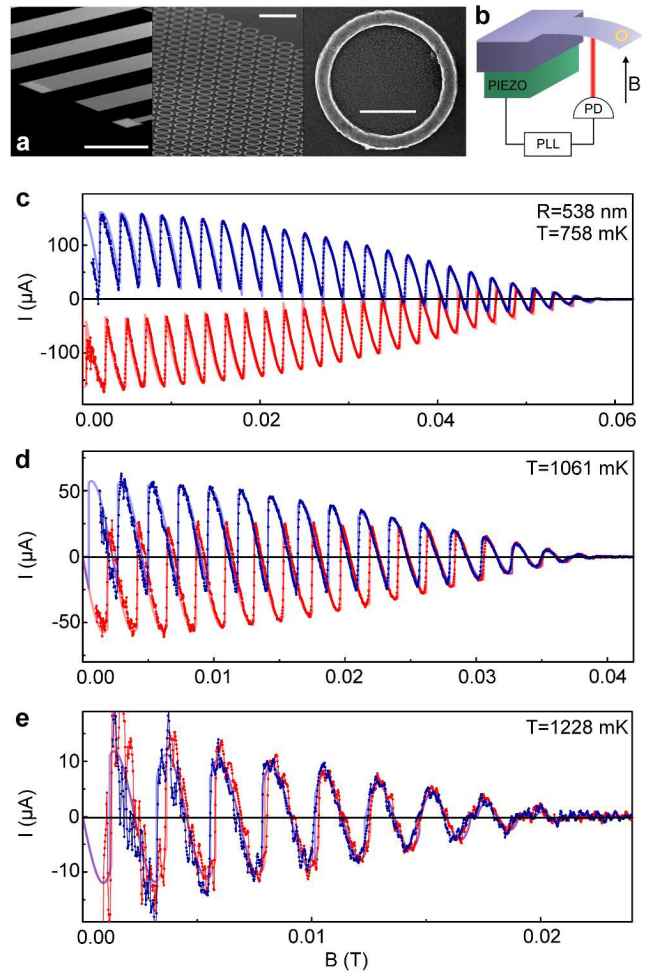,width=85mm}}}
\caption{ \textbf{Measured $\bm{I(B)}$.} (\textbf{a}) SEM photos of the sample. Left to right: several free-standing cantilevers, an array of rings on a single cantilever, a single Al ring from the array. The scale bars left to right are: 100 $\mu$m, 2 $\mu$m, 500 nm. (\textbf{b}) Measurement setup. A cantilever supporting rings is placed in a perpendicular magnetic field $B$. The cantilever's position is monitored by a laser interferometer (red). The signal from the photodiode (PD) is sent to a phase-locked loop (PLL) which drives a piezoelectric element (green) under the cantilever. The current in the rings is determined from the frequency of the PLL drive.
 (\textbf{c})-(\textbf{e}). Supercurrent per ring $I$ as function of magnetic field $B$ for rings with radius $R = 538$ nm at different temperatures $T$ (marked on each panel). Points are data; thick curves are the fits described in the text.
Red (blue) corresponds to increasing (decreasing) $B$. \break
}	
\label{current_one}
\end{figure}

\noindent \textbf{Description of the system}.  In this experiment four separate samples were measured. Each sample consists of an array of $100-1000$ nominally identical aluminum rings. Arrays were used to get a better signal-to-noise ratio. Ring radii of the four samples are $R=288$ to 780 nm, with nominal widths of $w=65$ to 80 nm and thickness $d=90$ nm. Detailed sample properties are listed in the Methods section and in the Supplementary Table 1. SEM photos of the sample are shown in Figure \ref{current_one}a.

The measurement setup is shown in Figure \ref{current_one}b.
A uniform magnetic field of magnitude $B$ is applied normal to the rings' equilibrium orientation. As the cantilever oscillates, current circulating in the rings experiences a torque gradient, which shifts the cantilever's resonant frequency by an amount $df$,  monitored by driving the cantilever in a phase-locked loop.
More details on the measurement setup are given elsewhere \cite{manuel,will_thesis}. In the configuration used here, $df= \kappa \, I  \, \Phi$, where $\Phi=B\, \pi R^2$ and $\kappa$ is a constant depending on the cantilever parameters, inversely proportional to the spring constant  \cite{ania_science,will_thesis}. A detailed description of the conversion of data from $df$ to $I$ is given in the Supplementary Notes 1 and 2 and Supplementary Figure 1.
	
\vspace{3mm}
\noindent \textbf{Metastable states and hysteresis}.
A superconducting ring is considered one-dimensional if its lateral dimensions are smaller than the coherence length $\xi$ and the penetration depth $\lambda$. The equilibrium properties of such a ring have three distinct temperature regimes, which are set by $R/\xi(T)$. For temperature $T$ only slightly below $T_{\mathrm{c}}$ such that $2R<\xi$, the ring is in a superconducting state for some values of $\Phi$ while for the other values it is in the normal state \cite{little_parks,little_parks2}, due to competition between the superconducting condensation energy and the flux-imposed kinetic energy of the supercurrent.
At slightly lower $T$ ($\xi<2R<\sqrt{3}\,\xi$), the condensation energy is slightly larger and for each value of $\Phi$ the ring has exactly one superconducting state. Finally, at even lower $T$ such that $2R>\sqrt{3}\,\xi$, the condensation energy is high enough to allow for several equilibrium states at a given $\Phi$. Depending on the ring's circumference, these three regimes may occur in the vicinity of  $T_{\mathrm{c}}$ described by the GL theory or may extend to lower temperatures, prompting the use of the empirical two-fluid model along with GL.

Figure \ref{current_one}c-e shows $I(B)$ for the sample with $R=538$ nm as $T$ is varied.
The red points show measurements taken while $B$ is increasing, and the blue points while $B$ is decreasing.
All the measurements exhibit sawtooth-like oscillations whose period is inversely proportional to the ring area $\pi R^2$.
The smooth parts of the sawtooth represent current $I_n$ in equilibrium states characterized by the order parameter winding number $n$, and the jumps correspond to phase slips between these states.
The jumps occur with flux spacing equal to the superconducting flux quantum $\Phi_0=h/2e$, indicating that $n$ changes by unity at each jump. Measured $I(B)$ curves for all other temperatures and ring sizes are given in Supplementary Figure 2.
The three qualitative regimes described previously are accessed by varying either $T$ or $B$, since
 they both diminish the condensation energy.
For low $T$ and $B$ the data are hysteretic, indicating the presence of multiple equilibrium states.
At sufficiently high $T$ or $B$ the hysteresis vanishes, indicating that only one superconducting state is available.
For the highest values of $B$ and $T$ there are ranges of $B$ over which $I=0$ (to within the resolution of the measurement), corresponding to the rings' re-entry into the normal state. In this so-called Little-Parks regime we observe the expected features: the persistent current goes through zero when the flux bias equals an integer number of flux quanta, whereas the winding number changes at half-integer values \cite{little_parks,little_parks2}. This is described in more detail in Supplementary Note 3 and shown in Supplementary Figures 3 and 4.

\vspace{3mm}
\noindent \textbf{Fit to theory}.
To compare these measurements with theory, we first identify the winding number $n$ of each smooth portion of $I(B)$. Then we simultaneously fit all of the smooth portions of $I(B)$ using the analytic expression derived from the GL theory for one-dimensional rings \cite{zhang}. This expression includes the rings' finite width $w$, which accounts for the magnetic field penetration into the ring volume and is crucial for reproducing the overall decay of $I$ at large $B$.
At each value of $T$, the fitting parameters are $\xi$ and the Pearl penetration depth $\lambda_{\mathrm{P}}=\lambda^2/d$,  appropriate when the bulk penetration depth $\lambda>d$ \cite{pearl,tinkham}, which holds.  The cantilever spring constant is assumed to be temperature independent, and is used as a global fit parameter for each sample, along with  the ring dimensions $w$ and $R$.
The resulting fits are shown as thick curves in Figure \ref{current_one}c-e.
The full set of fits to measured $I(B)$ for all $R$ and $T$ is shown in  Supplementary Figures 5 and 6, along with a more detailed description of the fitting procedure given in  Supplementary Note 4.

\begin{figure}[h!]
\centerline{\hbox{
\epsfig{figure=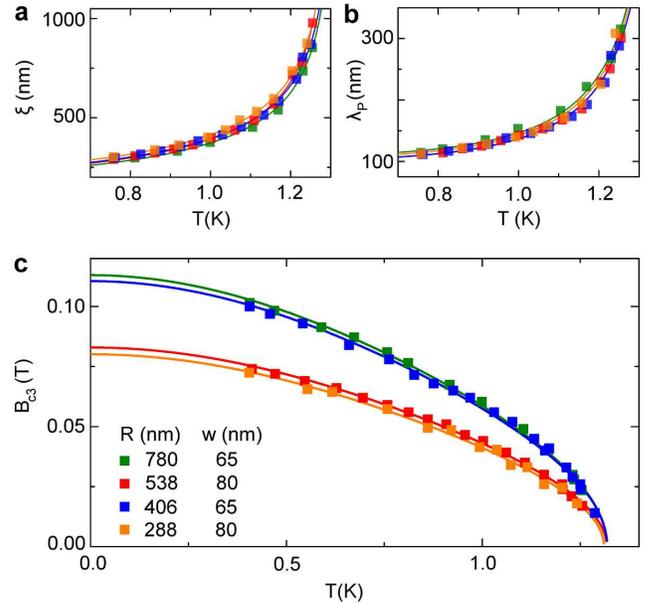,width=85mm}}}
\caption{\textbf{Coherence length, penetration depth and rings' critical field.} (\textbf{a})-(\textbf{b})  Coherence length $\xi$ and Pearl penetration depth $\lambda_{\mathrm{P}}$ as function of temperature.
The squares are the best-fit values from the GL fits described in the text. (\textbf{c})  Rings' critical field $B_{\mathrm{c} 3}$ as function of temperature. The squares are determined from measurements of $I(B)$. The lines in all panels are the fits described in the text. \break
}
\label{bc3}
\end{figure}

In each data set we identify the rings' critical field $B_{\mathrm{c}3}$, which we take to be the value of $B$ at which $I$ becomes indistinguishable from $0$ and remains so for all $B>B_{\mathrm{c}3}$. Note that the identification of $B_{\mathrm{c}3}$ is independent of any theoretical model. Next, we use the GL result for one-dimensional rings $B_{\mathrm{c}3}=3.67 \Phi_0/(2 \pi w \xi(T))$ \cite{vodolazov} to extract $\xi(T)$ (the fit parameters are $B_{\mathrm{c}3,0}\equiv 3.67 \Phi_0/(2 \pi w \xi_0)$ for each sample and $T_{\mathrm{c}}$ common to all the samples). The coherence lengths $\xi(T)$ extracted from the fits of $I(B)$ and from the $B_{\mathrm{c}3}(T)$ data agree with each other in the entire temperature interval and are approximated remarkably well by $\xi(T) = \xi_0\sqrt{(1+t^2)/(1-t^2)}$, where $t=T/T_{\mathrm{c}}$. The same relation inspired by the two-fluid model~\cite{tinkham} was used successfully to treat the thin-film upper critical field~\cite{tedrow_meservey,maloney}.
Along with $\xi(T)$, fits of $I(B)$ yield the temperature dependence of the Pearl penetration depth, which agrees well with the two-fluid model, $\lambda_{\mathrm{P}}(T) = \lambda_{\mathrm{P}0}/(1-t^4)$.
Figure \ref{bc3} shows the best-fit parameters $\xi$ and $\lambda_{\mathrm{P}}$, as well as $B_{\rm c3}$, all as function of $T$.
The best-fit values of $\xi_0$ ($\sim 200$ nm), $\lambda_{P0}$ ($\sim 100$ nm), $T_\mathrm{c}$ ($\sim 1.32$ K) and $B_{\mathrm{c}3,0}$, along with more details, are given in Supplementary Note 5 and Supplementary Table 1.
Lastly, we note that $B_{\mathrm{c}3}(T)$ should be independent of $R$ and proportional to $1/w$, consistent with the data in Fig. \ref{bc3}c.

\vspace{3mm}
\noindent \textbf{Criterion for deterministic phase slip}.
Figure \ref{current_one}c-e shows that on each branch $I_n$, the values of current at which the phase slips occur for increasing and decreasing $B$ are located nearly symmetrically around zero current.
To examine the locations of these phase slips quantitatively, we define $\Delta\phi_n^{\pm}=\phi_n^{\pm}-\phi_{\mathrm{min},n}$. Here $\phi_n^{\pm}$ is the experimental value of the normalized flux $\phi = \Phi/\Phi_0$ at which the transition $n \rightleftarrows n \pm 1$ occurs, and $\phi_{\mathrm{min},n}$ is the value of $\phi$ at which $I_n$ reaches zero. Flux $\phi_{\mathrm{min},n}$ is either directly measured, or obtained by extrapolation between sweep-up and sweep-down branches.  As defined, $\Delta\phi_n^+$ are positive (increasing $B$, for which $n\rightarrow n+1$) and $\Delta\phi_n^-$ are negative (decreasing $B$, for which $n\rightarrow n-1$).  (In the following we normalize all flux values by $\Phi_0$ and denote them by the character $\phi$.)

Our next step is to compare the experimental values of switching flux $\Delta\phi_n^{\pm}$ with theory. In the Langer-Ambegaokar picture, valid for a current-biased wire much longer than $\xi$, the barrier between states $n$ and $n-1$ vanishes when the bias current reaches the critical current $I_\mathrm{c}$ \cite{langer_ambegaokar}. In the case of a flux-biased ring, still for $R\gg \xi$, the barrier between states $n$ and $n\pm 1$ goes to zero at flux values

\vspace{-6mm}

\begin{equation}
\phi_{\mathrm{c},n}^{\pm}=\phi_{\mathrm{min},n}\pm\frac{R}{\sqrt{3}\,\xi}+O\left(\left(\frac{w}{R}\right)^2\right),
\label{lala}
\end{equation}

\noindent where $\phi_{\mathrm{min},n}=\frac{n}{1+\left(\frac{w}{2R}\right)^2}$. In the case $R\gtrsim\xi$, which corresponds to our experimental situation, it was shown that the system remains stable beyond $\phi_{\mathrm{c},n}^{\pm}$ and loses stability at a flux \cite{kramer_zimmermann,tuckerman}

\vspace{-4mm}

\begin{equation}
\phi_{\mathrm{f},n}^{\pm}=\phi_{\mathrm{min},n}\pm\frac{R}{\sqrt{3}\,\xi}\sqrt{1+\frac{\xi^2}{2R^2}}+O\left(\left(\frac{w}{R}\right)^2\right).
\label{kz}
\end{equation}

\noindent From these expressions we see that the switching flux is set by the ratio $R/\xi$ and therefore the precise determination of $\xi$ is crucial for quantitative comparison with theory.
To simplify this comparison it is convenient to refer all quantities not to zero field, but to the zero current field of each winding number, so  we define $\Delta\phi_{\mathrm{c},n}^{\pm}=\phi_{\mathrm{c},n}^{\pm}-\phi_{\mathrm{min},n}$ and $\Delta\phi_{\mathrm{f},n}^{\pm}=\phi_{\mathrm{f},n}^{\pm} -\phi_{\mathrm{min},n}$. Additional details on the free energy landscape close to the phase slip points are given in  Supplementary Note 6 and Supplementary Figure 7.

\begin{figure}[h!]
\vspace{2mm} \centerline{\hbox{
\epsfig{figure=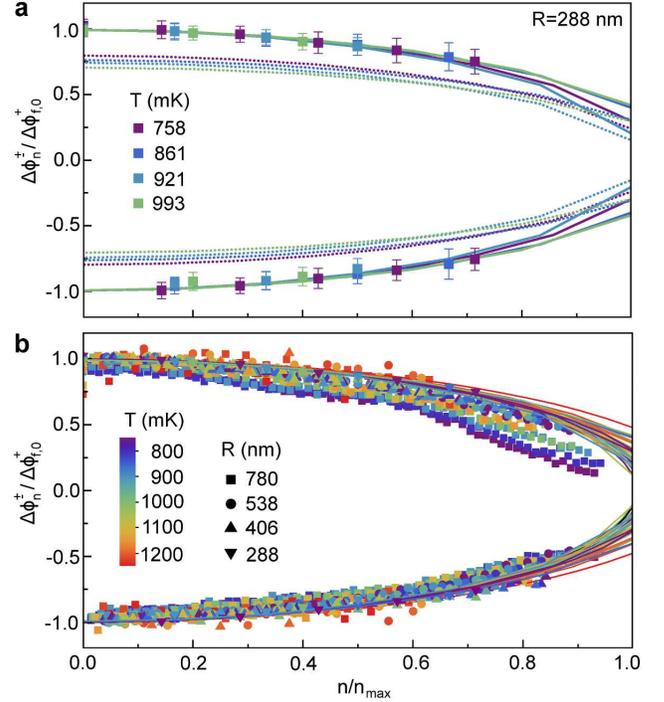,width=85mm}}}
\caption{\textbf{Phase slip flux as function of winding number.} Dots: experimental values; bars in  \textbf{a}: observed width of each jump due to size inhomogeneities in the array; full lines: prediction for the phase slip flux $\Delta\phi_{\mathrm{f},n}^{\pm}$; dotted lines in \textbf{a}: prediction for the phase slip flux $\Delta\phi_{\mathrm{c},n}^{\pm}$ (see text). Colors represent temperature. Panel (\textbf{a}) is the sample with $R=288$ nm and panel (\textbf{b}) contains data from all the samples. The normalization of the axes is explained in the text. \break
}
\label{switching}
\end{figure}

Figure \ref{switching} shows the measured $\Delta\phi_n^{\pm}$ as function of $n$.
The vertical axis in Fig. \ref{switching} is normalized to $\Delta\phi_{\mathrm{f},0}^+$.
The horizontal axis is normalized to the experimentally observed maximum winding number $n_{\mathrm{max}}$, where $n_{\mathrm{max}}\approx \frac{\sqrt{3}R^2}{w \xi}$. The ratio $n/n_{\mathrm{max}}$ is very close to $B/B_{\mathrm{c}3}$. There is a symmetry $\Delta\phi_n^+=-\Delta\phi_{-n}^{\:-}$ for $-n_\mathrm{max} \leq n \leq  n_\mathrm{max}$ so it suffices to consider $n \geq 0$.
Fig. \ref{switching}a shows the data for $R=288$ nm. The bars represent the width of the steep portion of the sawtooth oscillations, primarily due to the small size inhomogeneities in the array (see Supplementary Figures 8 and 9, and Supplementary Notes 7 and 8). In Fig. \ref{switching}b we show the data for all four samples, normalized such that all the data collapse together.
Supplementary Figure 8 shows the same data separated into four panels by ring size for a more detailed comparison.

The solid lines in Figure \ref{switching} show the predicted $\Delta\phi_{\mathrm{f},n}^{\pm}/\Delta\phi_{\mathrm{f},0}^+$ (see Eq. (\ref{kz})), whereas dotted lines in Figure \ref{switching}a show $\Delta\phi_{\mathrm{c},n}^{\pm}/\Delta\phi_{\mathrm{f},0}^+$ (Eq. (\ref{lala})).
The difference between the solid and dotted lines increases with the ratio $\xi(T)/R$ and is therefore the most pronounced for small rings (Fig. \ref{switching}a) or at high temperature due to the increase of $\xi(T)$.
We see that the prediction $\Delta\phi_{\mathrm{f},n}^{\pm}/\Delta\phi_{\mathrm{f},0}^+$, which includes the finite-circumference effect ($R\gtrsim \xi$), agrees well with the measured switching locations over the full range of $T$, $B$, and $R$.

\begin{figure}[h!]
\centerline{\hbox{
\epsfig{figure=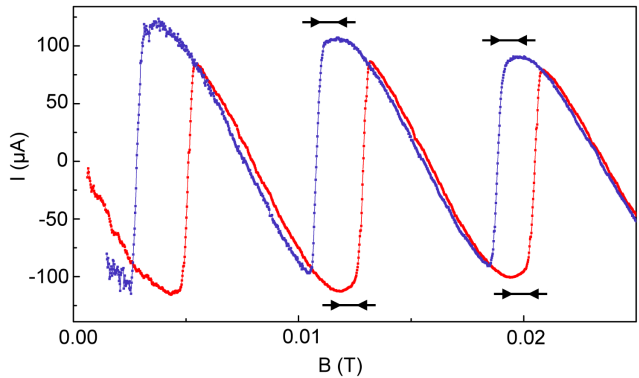,width=85mm}}}
\caption{\textbf{Direct observation of the finite-length correction to the phase slip criterion.} Supercurrent per ring $I$ as function of magnetic field $B$ for rings with radius $R=288$ nm and temperature $T=861$ mK.
Red (blue) points: increasing (decreasing) $B$.
The regions over which $I(B)$ diminishes at fixed winding number are indicated by black arrows. Diminishing of current after having reached a maximum but before the phase slip event is due to the finite-length correction to the phase slip criterion. \break }
\label{pudding}
\end{figure}

The finite-circumference effect can also be seen directly in Fig. \ref{pudding}, which shows $I(B)$ over a narrow range of $B$ for the smallest rings. For both increasing $B$ (red) and decreasing $B$ (blue) each sawtooth oscillation reaches a maximum current and then starts to diminish before the switching occurs, as seen in the regions indicated by the black arrows.

\vspace{3mm}
\noindent \textbf{Damping}.
For $T$ well below $T_{\mathrm{c}}$, once $\xi$ is exceeded sufficiently by the circumference of the ring,
there are typically multiple free energy minima into which the system may relax. Despite this freedom, we find that the winding number always changes as $|\Delta n| = 1$. This is seen for all measured rings
and all $T$ down to the lowest value $T=460$ mK. In contrast, previous experiments~\cite{pedersen,vodolazov} with Al rings at $T<400$ mK
have found $|\Delta n| > 1$.

We expect the tendency for $|\Delta n| > 1$ to increase with lowering $T$. Indeed, a circulating current of almost-critical value and temperature $T$ close to $T_\mathrm{c}$ result, respectively, in the suppression of the BCS singularity in the electron density of states and high density of Bogoliubov quasiparticles in a superconductor~\cite{tinkham}. These are the two conditions making the dynamics of the order parameter dissipative and well-described~\cite{levchenko-kamenev} by the time-dependent Ginzburg-Landau equation (TDGL). In the context of phase slips~\cite{mccumber_halperin} it determines a viscous motion of the phase difference across the phase slip, $\varphi(\tau)$ ($\tau$ being time), down the monotonic part of the effective potential relief $V(\varphi)$, and this viscous motion results in $|\Delta n| = 1$. In the opposite limit of low temperatures, the quasiparticle density is low, and we may try considering the phase slip dynamics in terms of the Andreev levels associated with the phase slip. Their time evolution caused by the variation of $\varphi(\tau)$ results in Landau-Zener tunneling between the occupied and empty levels, thus leading to dissipation~\cite{averin} of the kinetic energy of the condensate (the energy is irreversibly spent on the production of quasiparticles). Our estimate (Supplementary Note 9) of the energy lost in this way is $E_{\rm diss}\sim (\hbar S/e^2\rho\xi)\Delta$, where $\rho$ and $S$ are, respectively, the normal state resistivity and cross-section of the aluminum wire forming the ring, and $\Delta$ is the superconducting gap; a numerical proportionality factor is beyond the accuracy of the estimate.

The condensate energy difference between the two metastable states involved in a $|\Delta n| = 1$ transition is $E_{\Delta n=1}=(\hbar/e)j_\mathrm{c} S\sim (\hbar S/e^2\rho\xi)\Delta$; here $j_\mathrm{c}\sim\Delta/(e\rho\xi)$ is the critical current density. Furthermore, the lower of the two states is protected by a barrier $\delta F_{\Delta n=1}\sim (\xi/R)^{5/2}E_{\Delta n=1}$ (the estimate is easily obtained from the Langer-Ambegaokar~\cite{langer_ambegaokar} scaling, $\delta F \propto (1-j/j_\mathrm{c})^{5/4}$, of the barrier with the current density $j$, see Supplementary Note 9). The height of the barrier is smaller for larger rings.

We find the irreversibly lost energy $E_{\rm diss}$ to be of the order of the energy difference between the two metastable states $E_{\Delta n=1}$. The above estimates​, given their limited accuracy, allow (but do not guarantee) the condensate to have a sufficient excess of kinetic energy to overcome a small barrier  out of  the metastable state with $\Delta n=1$. In addition to higher temperatures, in a notable difference from the previous experiments the rings studied here had smaller $R$, providing a better protection of the metastable states.

\vspace{3mm}
\noindent \textbf{\large{Discussion}}

\noindent We have studied the persistent current in arrays of flux-biased uniform one-dimensional superconducting Al rings. We found detailed agreement with GL theory, including the location of deterministic phase slips, which are predicted to occur when the barrier confining the metastable state occupied by the ring goes to zero. In one dimension GL theory has a relatively simple, analytic form, and due to their small width, our rings are strictly in the one-dimensional limit, in contrast to those studied previously \cite{vodolazov,pedersen,bluhm_moler}. As a result, GL theory provides detailed knowledge of the free energy landscape in these samples. This should enable  systematic study of thermal and quantum phase slips in isolated rings, and progress towards the quantitative understanding of coherent quantum phase slips \cite{astafiev1,astafiev2}, one of the outstanding goals in the field \cite{buchler,mooij_nazarov,mooij_harmans}.

\vspace{3mm}
\noindent \textbf{\large{Methods}}

\small{
\noindent \textbf{Sample fabrication.}
Ring radii of the four measured samples are $R=288, 406, 538$ and 780 nm, nominal widths are $w=65$ nm (for $R=406, 780$ nm) and 80 nm (for $R=288, 538$ nm) and thickness $d=90$ nm. Further details on sample properties are listed in the Supplementary Table 1. Each array is fabricated on a Si cantilever of length $\sim 400\; \mu$m, thickness 100 nm and width $\sim 60 \;\mu$m, with resonant frequency $f \sim 2$ kHz, spring constant $k \sim 1$ mN m$^{-1}$ and quality factor $ Q \sim 10^5$.
Cantilevers are fabricated out of a silicon-on-insulator wafer. They are patterned out of the top silicon layer by means of optical lithography followed by a reactive ion etch. Rings are then fabricated on top of patterned cantilevers using standard e-beam lithography with a PMMA mask, into which Al is evaporated in a high-vacuum thermal evaporator. After lift-off, the top of the wafer is protected and the backing silicon layer is etched in KOH, followed by a BOE etch of the SiO$_2$ layer and drying in a critical point dryer. This results in cantilevers being fully suspended. Further details on the the fabrication process are given elsewhere \cite{ania_science,will_thesis}.

\small{
\noindent \textbf{Data availability}.
The data that support the findings of this study
are available from the corresponding author upon request.}

\vspace{5mm}

\newpage

\noindent \textbf{Author contributions}

\small{\noindent A.L. and I.P performed the measurement. All authors conducted the analysis. I.P., J.H. and L.G. wrote the manuscript. All authors discussed the results and commented on the manuscript. }

\vspace{3mm}

\begin{acknowledgments}
\noindent \textbf{Acknowledgments}

\small{
\noindent We thank Amnon Aharony, Richard Brierley, Michel Devoret, Ora Entin-Wohlman, Alex Kamenev, Konrad Lehnert, Hendrik Meier and Zoran Radovi\'{c}  and for useful discussions, and Ania Jayich and Will Shanks for fabricating the samples. We acknowledge support from the National Science Foundation (NSF) Grant No. 1106110 and the US-Israel Binational Science Foundation (BSF). L.G. was supported by DOE contract DEFG02-08ER46482.}
\end{acknowledgments}

\clearpage

\onecolumngrid


\normalsize

\renewcommand{\figurename}{Supplementary Figure}
\renewcommand{\tablename}{Supplementary Table}

\setcounter{figure}{0}
\setcounter{equation}{0}

\subsection{Supplementary Note 1. Background removal}

We measure the shift of the resonant frequency of the cantilever $f$ as function of field $B$. An example of raw data taken for a ring with radius $R=406$ nm at temperature $T=762$ mK is shown in Supplementary Figure \ref{data_t}a. Red trace corresponds to the sweep up of bias field and blue to the sweep down. In addition to the sawtooth oscillations associated with the rings' superconductivity, we observe that $f$ also undergoes a small drift as a function of time and of $B$. To remove this background, we fit the data above the rings' critical field to a third-order polynomial, which is shown as the black curve in Supplementary Figure \ref{data_t}a. We subtract this fit from $f$ to obtain the frequency shift $df$ due to the magnetic moment of the rings (Supplementary Figure \ref{data_t}b).

\begin{figure}[h]
 \centerline{\hbox{
\epsfig{figure=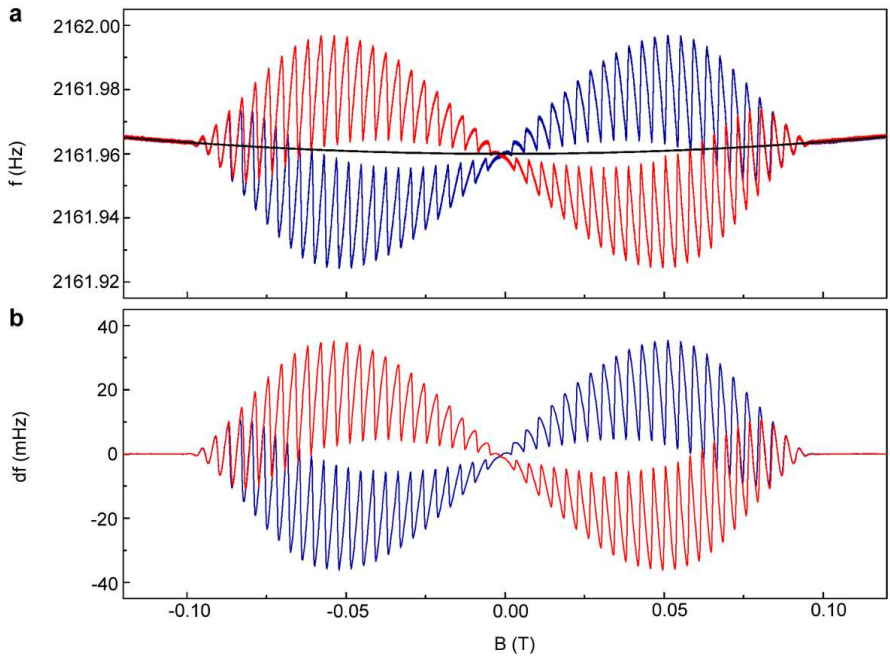,width=120mm}}
}
\caption{\textbf{Background removal.} (\textbf{a})  Raw data, cantilever frequency shift as function of field for increasing $B$ (red) and decreasing $B$ (blue) for the rings with radius $R=406$ nm and $T=762$ mK. Third order polynomial background is shown as the black curve. (\textbf{b}) Cantilever frequency shift after background substraction and averaging. The signal is due to the magnetic moment of the supercurrent. \break }
\label{data_t}
\end{figure}

\subsection{Supplementary Note 2. Supercurrent as function of field}

\begin{figure}[p]
\centerline{\hbox{
\epsfig{figure=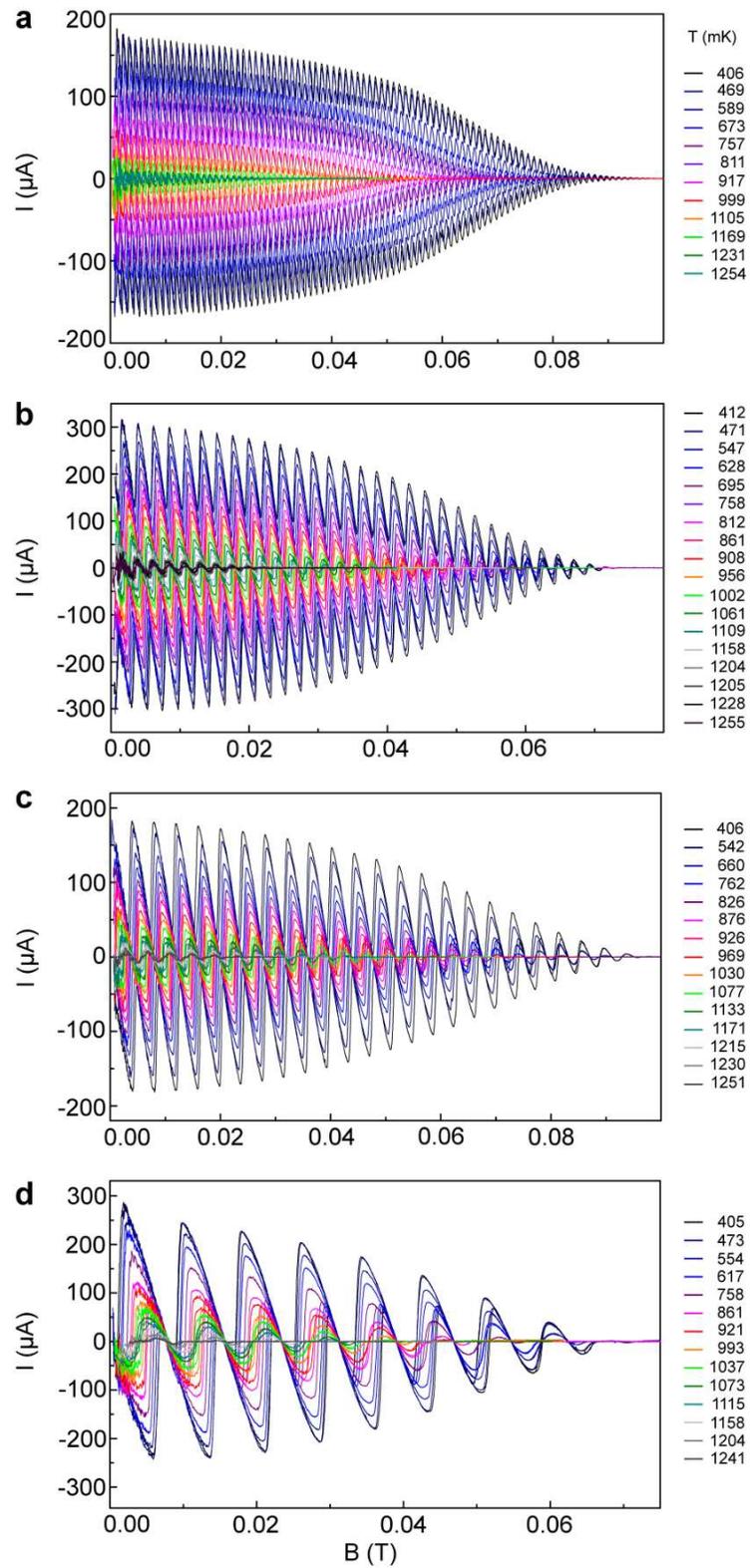,width=100mm}}}
\captionsetup{justification=raggedright}
\caption{\textbf{$\,$ Measured supercurrent as function of field.} $\,$ Shown  are different samples with  $R=780,\:538,\:406,\:288$ nm ((\textbf{a}) to (\textbf{d})), in the full temperature range.
Lower curves of the same color correspond to increasing $B$ and upper ones to decreasing $B$. \break }
\label{ib_cl}
\end{figure}

In our measurement configuration  the magnetic field is perpendicular to the rings' surface and the frequency shift is related to supercurrent $I$ as $df(B)=\kappa \, I(B)\, B R^2 \pi$, where $\kappa$ is a cantilever-specific constant \cite{ania_science,wills_thesis}.  This constant depends on the cantilever's resonant frequency, spring constant, length and the number of rings on it. The resonant frequency is measured in the phase-locked loop, the length is measured by optical imaging, and the number of rings is known from the lithography pattern. The spring constant $k$ is obtained as a fitting parameter of the Ginzburg-Landau fit, as explained in the main text and here in the following section. The best-fit value is within 20$\%$ of the nominal value computed as $k=(2\pi \!f)^2 m_{\rm eff}$, where $m_{\rm eff} = m/4$ is the effective mass of the cantilever and $m$ is the cantilever's actual mass. The rings' radius is also obtained from the Ginzburg-Landau fits. It is highly constrained by the period of Aharonov-Bohm oscillations, with the result that the statistical error on the best-fit value is $\approx 1$ nm. The values returned by this fit agree well with the values measured by SEM observations.

The supercurrent obtained from the frequency shift as $I(B)=df(B)/( 2 \pi \kappa R^2 B)$ is shown in Supplementary Figure \ref{ib_cl}. Every panel shows data measured on a sample with a different ring size in the full available temperature range.

The signal becomes very noisy close to zero field, and is not displayed for $B$ very close to 0. This is because $I \propto df/B$ and for fields close to zero, dividing the signal $df$ by $B$ leads to unreliable results.

\subsection{Supplementary Note 3. Little-Parks regime}

As mentioned in the main text, at high $T$ and high $B$ the rings exhibit the Little-Parks effect: as function of bias flux the rings alternate between the superconducting and the normal state due to the competition between the superconducting condensation energy and the kinetic energy of the current imposed by bias flux \cite{little_parks,little_parks2}. This can be seen directly in the $I(B)$ curves close to the rings' critical field, as they show regions of zero current (normal state) between regions of non-zero current (superconducting state). This is illustrated in Supplementary Figure \ref{lp}, where the normal state regions are denoted by black arrows.

\begin{figure}[h]
\centerline{\hbox{
\epsfig{figure=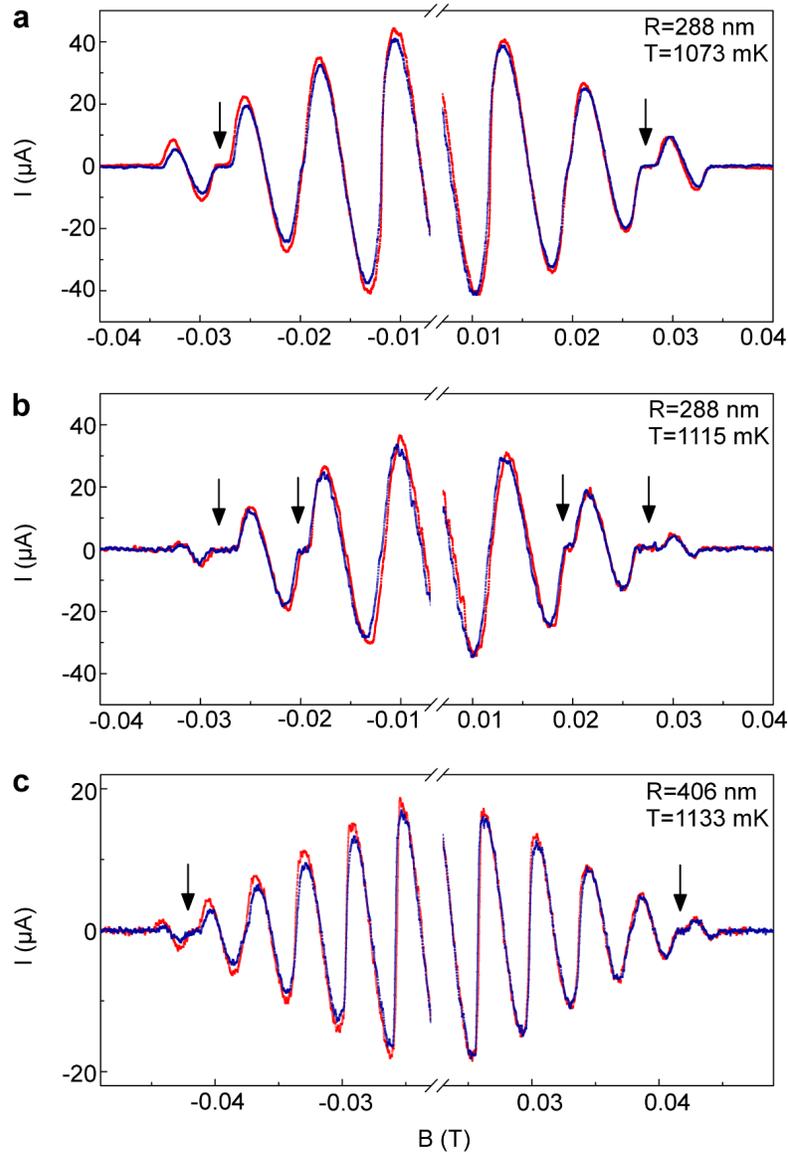,width=105mm}}}
\captionsetup{justification=raggedright}
\caption{\textbf{Little-Parks regime.} Measured supercurrent $I(B)$ for ring sizes and temperatures marked in each panel. Normal state regions
 $(I=0)\,$  are denoted by black arrows. Red curves correspond to sweep up and blue to sweep down. \break  }
\label{lp}
\end{figure}

It is known from the Little-Parks effect that the supercurrent velocity, and therefore also supercurrent, reaches zero when bias flux $\Phi_\mathrm{min}=n \Phi_0$, where $n$ is an integer \cite{little_parks,little_parks2,tinkham}. GL theory shows this is the case not only in the Little-Parks region, but in the full field region. Due to finite ring width there is a small correction on this condition \cite{zhang} and in fact supercurrent is zero when

\vspace{-5mm}

\begin{equation}
\Phi_\mathrm{min}=\frac{n \Phi_0}{1+\left(\frac{w}{2R}\right)^2}.
\label{min}
\end{equation}

\noindent The correction $(w/2R)^2$ is very small in our experiment (a few percent), but at high winding number $n$ it may lead to an observable deviation from the integer value. This is shown in Supplementary Figure \ref{qpz} where full squares are the measured flux values at which the current goes to zero and the full black curve in Supplementary Figure \ref{qpz}a is the theoretical prediction (Supplementary Equation (\ref{min})), using parameters $w$ and $R$ obtained by the GL fit. In Supplementary Figure \ref{qpz}b a linear background $n \Phi_0$ has been subtracted from the data and from the theoretical prediction, and one directly sees the small linear slope due to the finite-width correction. At this  scale we can distinguish red (sweep up) from blue (sweep down) data. The slight discrepancy between data and theory probably reflects the imperfect background subtraction in the $df(B)$ data, causing  the observed zero current to deviate slightly from the actual value. Ideally, measured sweep up and sweep down values should coincide, so the scatter between them also points to noise in the readout.

\begin{figure}[h]
\centerline{\hbox{
\epsfig{figure=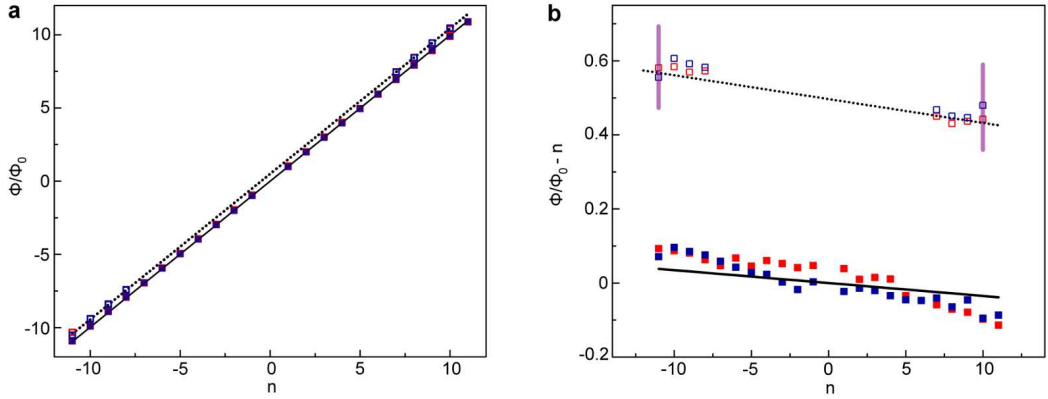,width=140mm}}}
\caption{\textbf{Quantization in the Little-Parks regime.} Full squares denote normalized flux values at which current goes through zero, and empty squares denote normalized flux at which the winding number changes by one, both as function of winding number $n$. Black lines are the theoretical predictions for both cases, full and dotted respectively. In panel (\textbf{b}) with respect to (\textbf{a}) the linear background $\Phi/\Phi_0=n$ is subtracted both from data and theoretical predictions, which enables to outline the trend due to finite-width correction. Purple vertical lines denote the normal-state regions. Red squares are for sweep up and blue for sweep down. Here $R=406$ nm and $T=1133$ mK.  \break }
\label{qpz}
\end{figure}

Also shown in Supplementary Figure \ref{qpz} with empty squares are measured flux values at which the winding number changes by one in the Little-Parks region. Those flux values are expected to be very close to $(n+1/2)\Phi_0$. Again there is a small correction of the order $(w/2R)^2$ which is too cumbersome to write explicitly.  The black dotted line is the theoretical prediction including this correction. Supplementary Figure \ref{qpz}b shows this same data and theory as Supplementary Figure \ref{qpz}a, with the linear contribution $\Phi=n \Phi_0$ subtracted. Vertical purple lines show the extent of the normal regions (denoted by black arrows in Supplementary Figure \ref{lp}).

\subsection{Supplementary Note 4. Ginzburg-Landau fit for a one dimensional ring with finite width}

We fit the data using a theory which includes the effects of finite ring width. More specifically, we use the expression for supercurrent $I(B)$ as given in Eq. (7) of \cite{zhang}. That expression uses the relation of $I(B)$ to the coherence length $\xi$ and field penetration depth $\lambda_P$ prescribed by the Ginzburg-Landau theory, but does not specify the temperature dependence of $\xi$ and $\lambda_P$. The latter dependence is found from the fits to the frequency shift data $df$ taken in a broad range of temperatures and fields ($I(B)$ and $df$ are related to each other via the spring constant). For brevity, we will refer to that procedure as to the Ginzburg-Landau fit.

To perform the fit, we first identify the winding number of each segment of $df(B)$. For the measurements taken with increasing $B$, we count the number of segments (i.e., the regions of smoothly varying $df$ between jumps) between $B_{\rm c3}$ and $-B_{\rm c3}$. This number is $2 n_{\rm max}+1$, where $n_{\rm max}$ is the maximum winding number. We thus determine $n_{\rm max}$. Then we start from $B_{\rm c3}$ and count down from $n_{\rm max}$ to zero. We apply the equivalent process to measurements taken with decreasing $B$.

\begin{figure}[h]
 \centerline{\hbox{
\epsfig{figure=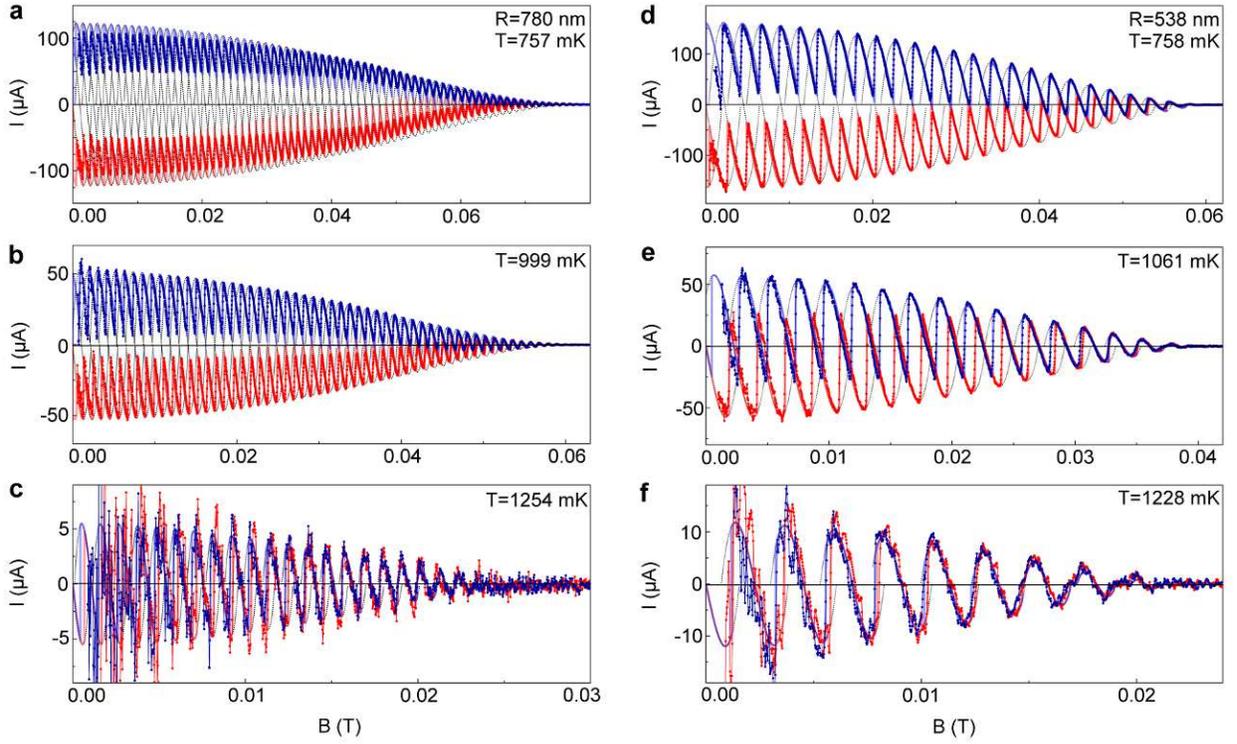,width=165mm}}
}
\caption{\textbf{Supercurrent vs. field and Ginzburg-Landau fit.} Data is shown for different ring sizes (columns) and temperatures (marked on each panel). Points and thin curves: data; thick curves: Ginzburg-Landau fit (see text). Red curves on each graph correspond to sweeping the field up, and the blue ones to sweeping down. Thin black dotted curves: the Ginzburg-Landau fit, extended over the full field range of each winding number.
\break }
\label{ifit1}
\end{figure}

\begin{figure}[h]
\vspace{2mm}
 \centerline{\hbox{
\epsfig{figure=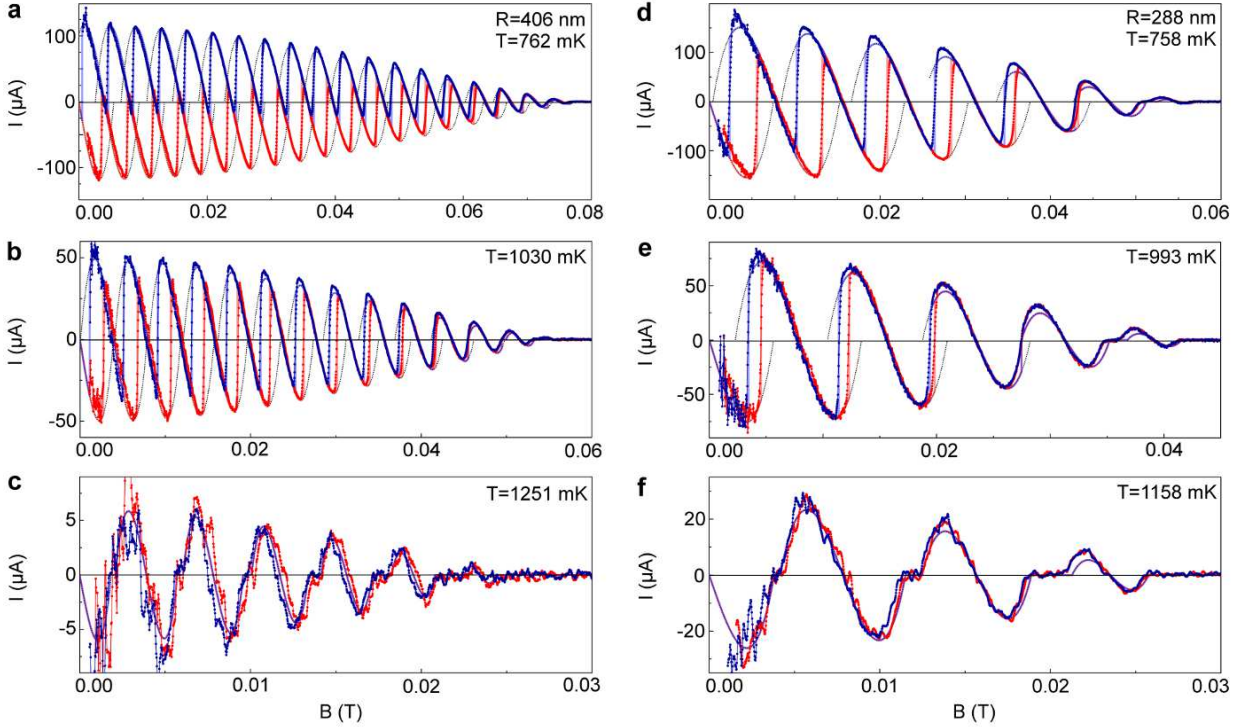,width=165mm}}
}
\caption{\textbf{Supercurrent vs. field and Ginzburg-Landau fit.} Data is shown for different ring sizes (columns) and temperatures (marked on each panel). Points and thin curves: data; thick curves: Ginzburg-Landau fit (see text). Red curves on each graph correspond to sweeping the field up, and the blue ones to sweeping down. Thin black dotted curves: the Ginzburg-Landau fit, extended over the full field range of each winding number.
\break }
\label{ifit2}
\end{figure}

As explained in the main text, it is a global fit which fits the entire $I(B)$ measurement (i.e., for all winding numbers and for $B$ increasing and decreasing). The fitting parameters are: superconducting coherence length $\xi$, penetration depth $\lambda$, ring radius $R$, ring width $w$ and  spring constant $k$. Of these, we expect $R$, $w$ and $k$ to be fixed for each sample (i.e., to not change with temperature), so we first undertake a preliminary fit to determine these three parameters.
In these preliminary fits, there is a degeneracy between $\lambda$ and $k$, since they both set the amplitude of the signal: $\lambda$ affects the condensation energy, and therefore the amplitude of the current, while $k$ affects the proportionality constant $\kappa$ between current and frequency shift. Therefore we first set the starting value $k_{\rm in}$ to its calculated nominal value (using the expression given in the Supplementary Note 2) and $\lambda_{\rm in}$ such that $B_{\rm c0}$ (the zero temperature bulk critical field, set by the product of $\xi_0$ and $\lambda_0$, where zeroes denote the zero temperature value) is 0.01 T, as expected for aluminium \cite{tinkham}. Then we run the fit for each of the $I(B)$ measurements (i.e., at different $T$) for that sample. We then fix $k$ to be the mean of the values returned by these preliminary fits. Values for $R$ and $w$ are fixed in the same way: by picking the mean of the values obtained from fits at different temperatures. The scatter between the obtained values for $k$, $R$ and $w$ at different temperatures is rather small (a few percent for $k$ and $w$ and less than 1 nm for $R$).

In the second round of the fit only two fitting parameters remain, $\xi$ and $\lambda_P$. Note that $\xi$ also affects the condensation energy, and therefore the amplitude of the signal, but it is not degenerate with $k$ and $\lambda_P$ since it is very accurately set by the rings' critical field $B_{\rm c3}\propto \xi^{-1}$, as detailed in the main text. This is in a sense lucky because our subsequent conclusions on the switching flux value hinge on the precise determination of $\xi$. This can be seen from Eqs. (1) and (2) in the main text which show that the switching flux criteria depend only on $R$ and $\xi$.  The temperature dependence $\xi(T)$ and $\lambda_P(T)$ found this way agrees well with the expectations based on theory and on earlier measurements of the thin-film critical field, see the main text.

We have made measurements for $T > 400$ mK. We have tried fitting below 750 mK ($\sim T_\mathrm{c}/2$) but we have found that the values of $w$, $\lambda_0$ and $\xi_0$ don't converge to a fixed value like they do for $T>750$ mK. This likely reflects the decreasing applicability of GL theory at lower temperatures.

The result of the measurement and the Ginzburg-Landau fit are shown in Supplementary Figures \ref{ifit1} and \ref{ifit2}, where data is shown as points connected by thin curves and the fit is shown as thick curves. Three temperatures spanning the whole measured range are shown for each ring size. Red curves on each panel are for sweep up and blue for sweep down. The dotted black curves show the fit results extended over the full  field range for each winding number; note that the portion of the dotted black curve occupied when $B$ is increasing (red) is different from the part occupied for when $B$ is decreasing (blue) in the hysteretic part of $I(B)$.

The most pronounced discrepancy between the data and the fit is found for the biggest ring at the lowest temperatures, see Supplementary Figure \ref{ifit1}a, where the ring's self-inductance $L$ starts to play a role. In this regime, we estimate $L I \sim 0.13\, \Phi_0$, which may lead to non-negligible skewing of the rings' current-phase relation \cite{fink_grunfeld}. For the rest of the measurements considered here, the effects of  $L$ are unimportant, i.e. $LI \ll \Phi_0$. For the smallest to largest ring size we have computed the expected $L = 0.5 - 2.3$ pH. This gives $L I \sim 0.03, 0.05,0.07,0.13 \,\Phi_0 $  respectively at $ 750$ mK. At higher temperatures $LI$ is less since  $I$ decreases with temperature.

\subsection{Supplementary Note 5. Sample parameters}

Sample parameters are listed in Supplementary Table \ref{tabela}.

\begin{table*}[h!]
\begin{tabular}{ |c|c|c|c|c|c|c|c|c|c|c| }
 \hline
  N$^o$ &  $ \; R_{ \rm nom}$ (nm)  & $ \; R_{\rm  GL}$ (nm )& $ \; w_{\rm  nom}$ (nm) & $ \; w_{\rm  GL}$ (nm)& $\quad  N \quad$  &  $ \; \xi_0$ (nm)  &  $ \; \lambda_0$ (nm)&  $ \; \lambda_{\mathrm{P}0}$ (nm)  &  $ \quad B_{\mathrm{c}3,0}$ (T) & $ \quad B_{\mathrm{c}3,0}^{\;\mathrm{GL}}$ (T)  \\
  \hline
  \hline
  1 & 250 & 288 & 80 & 65 &  1680 & 214(2)   &  97(1) &  104(2)  &  0.0796(6) & 0.087(1) \\
  \hline
  2 & 375 & 406 & 65 & 48 &  990 & 202(2)    &  95(1) &  100(2)  & 0.1107(7) & 0.125(1) \\
  \hline
  3 & 500 & 538 & 80 & 65 &  550 & 208(2)    &  95(1) &  101(2)  & 0.0830(6) & 0.089(1) \\
  \hline
  4 & 750 & 780 & 65 & 51 &  242 & 190(3)    &  98(1) &   107(2)  & 0.1131(7) & 0.125(2) \\
 \hline
\end{tabular}
\caption{\textbf{Summary of sample parameters.} For each sample the table gives the nominal lithographic ring radius $R_{\mathrm{nom}}$ and width $w_{\mathrm{nom}}$, as well as the values $R_{\mathrm{GL}}$ and $w_{\mathrm{GL}}$ obtained as global fit parameters. The number of rings on each cantilever is $N$. $\xi_0$, $\lambda_{\mathrm{P}0}$, and $B_{\mathrm{c}3,0}$ are the zero temperature values of the coherence length, Pearl penetration depth, and critical field $B_{\mathrm{c}3}$  determined from the fits in Fig. 2 in the main Text. The penetration depth $\lambda_0=\sqrt{\lambda_{\mathrm{P}0}\,d}$. $B_{\mathrm{c}3,0}^{\;\mathrm{GL}}$ is calculated using $\xi_0$ and $w_{\mathrm{GL}}$, as described in the text. The quoted error in the final digit of each fit value corresponds to the statistical uncertainty of the fit (one standard deviation). \break }
\label{tabela}
\end{table*}

The values of $\xi_0$ and $B_{\mathrm{c}3,0}$ from Figure 2a and c in the main text and given in the Supplementary Table \ref{tabela} can be compared against two separate estimates. First, we note that $\xi_0$ can also be determined via transport measurements, using the relationship $\xi_0 = 0.855 \sqrt{\xi_0^{b} l_e}$ \cite{tinkham}, where $\xi_0^{b}=1.6 \,\mu$m is the bulk Al coherence length and $l_e$ the electron mean free path.
Transport measurements of Al wires that were co-deposited with the rings studied here give $l_e = 35 \pm 5$ nm \cite{ania_science}; this corresponds to $\xi_0 = 205 \pm 15$ nm, in close agreement with the values inferred from the measurements of $I(B)$.
Second, we note that $B_{\mathrm{c}3,0}$ can be calculated directly from $B_{\mathrm{c}3,0}  = 3.67 \Phi_0/(2 \pi w \xi_0) $ using the values of $\xi_0$ determined from the fits in Figure 2a in the main text and $w_{\mathrm{GL}}$. The results of this approach are listed in Supplementary Table \ref{tabela} as $B_{\mathrm{c}3,0}^{\;\mathrm{GL}}$.
For each sample, $B_{\mathrm{c}3,0}^{\;\mathrm{GL}}$ and $B_{\mathrm{c}3,0}$ agree to $\approx 10 \%$.

Figure 2 in the main text shows that $\xi$ and $\lambda_{P}$ obtained as fitting parameters follow the expected empirical temperature dependence, while measured rings' critical field $B_{\mathrm{c3}}$ follows the Ginzburg-Landau prediction. The critical temperature is obtained as a fitting parameter from all three panels separately, and it is found to be: $1.316 \pm 0.001, 1.391 \pm 0.004$ and $1.318 \pm 0.002$ K respectively. We see that the critical temperature obtained from the fit of $\lambda_P$ is larger by 5 $\%$. This is possibly due to the fact that at low temperature the thickness of the rings is not much larger than $\lambda$ and the system is marginally in the regime where Pearl penetration depth applies.

\subsection{Supplementary Note 6. Free energy}

With values of $\xi(T)$ and $\lambda_{\mathrm{P}}(T)$ obtained by fitting $I(B)$, it is straightforward to calculate the free energy $F_n(B)$ of each equilibrium state \cite{zhang}. The black line in Supplementary Figure \ref{free_energy}a shows $F_n(B)$ for the rings with $R = 406$ nm and $T = 876$ mK. In Supplementary Figure \ref{free_energy}a, the red (blue) curves show the path taken by the rings as $B$ is increased (decreased). The path is determined by using the values of $n$ inferred from the data. Supercurrent is related to the free energy as $I \propto -\partial F/\partial B$.

\begin{figure}[h!]
\centerline{\hbox{
\epsfig{figure=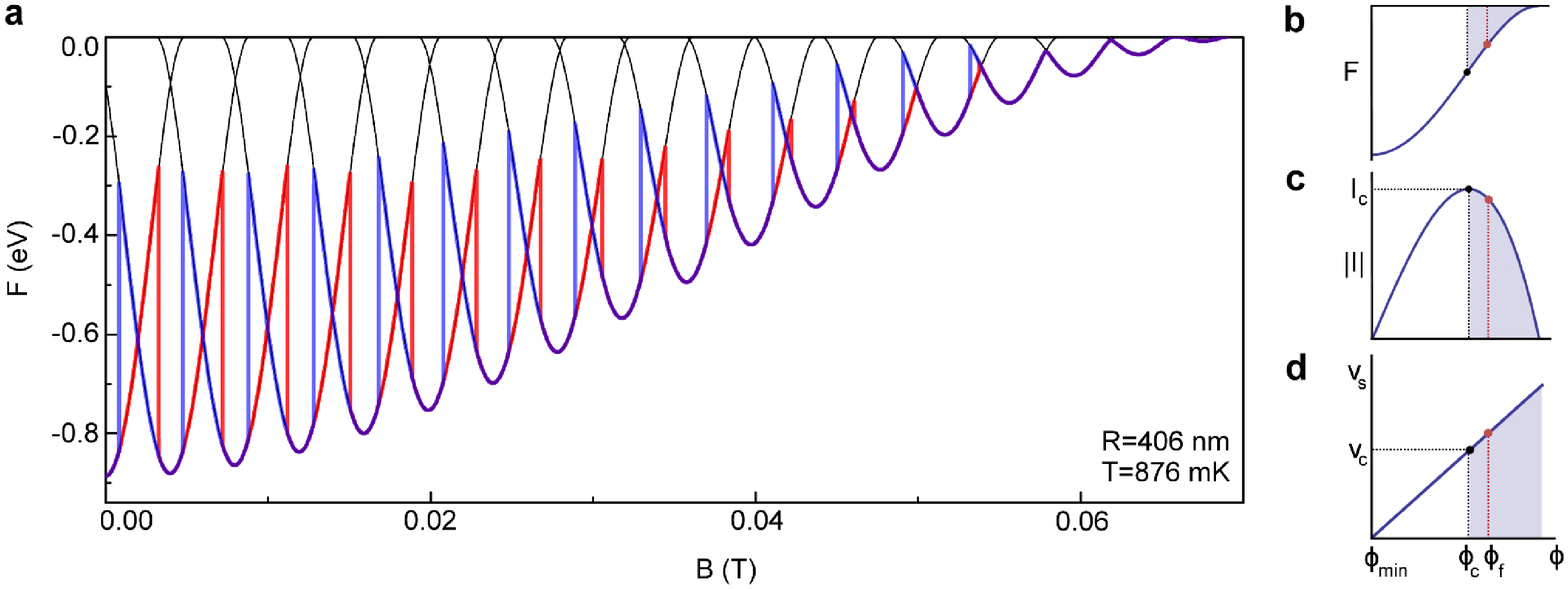,width=150mm}}} 
\caption{\textbf{Free energy.} (\textbf{a}) The black curve shows the equilibrium free energy $F_n$ as a function of $B$ for all $n$, for the ring with $R = 406$ nm at $T = 876$ mK. In this panel, the $F_n$ are calculated from the fit parameters. The red (blue) curve shows the path followed by the system as $B$ is increased (decreased). (\textbf{b}) Free energy of an equilibrium state as a function of flux. (\textbf{c}) Absolute value of supercurrent as a function of flux. (\textbf{d}) Velocity of the superconducting condensate as a function of flux. The black dots in panels \textbf{b}, \textbf{c} and \textbf{d} denote instability points $\phi_{\mathrm{c}}$, and red dots  $\phi_{\mathrm{f}}$ (see main text). Shaded regions are unstable in the sense specified in the text. \break }
\label{free_energy}
\end{figure}

Supplementary Figure \ref{free_energy}a shows that the phase slips for increasing and decreasing $B$ are located nearly symmetrically around the minima of $F_n(B)$. Closer inspection shows that the phase slips occur near the inflection points of $F_n(B)$. To examine the location of these phase slips quantitatively, in the main text we define $\Delta\phi_n^{\pm}=\phi_n^{\pm}-\phi_{\mathrm{min},n}$, where $\phi_n^{\pm}$ is the experimental value of the normalized flux $\phi = \Phi/\Phi_0$ at which the transition $n \rightleftarrows n \pm 1$ occurs, and $\phi_{\mathrm{min},n}$ is the value of $\phi$ at which $F_n$ reaches its minimum value (or, as stated in the main text, where the current $I_n \propto -\partial F_n/\partial \phi$ reaches zero, see also Supplementary Eq. (\ref{min})). As defined, $\Delta\phi_n^+$ are positive (increasing $B$) and $\Delta\phi_n^-$ are negative (decreasing $B$).

Both in a current-biased wire and in a flux-biased ring  the phase of the order parameter at equilibrium is $\phi=k s$, where $k$ is a wave-vector, and $s$ the longitudinal coordinate along the wire or ring. Supercurrent is then $I\propto k(1-k^2)$ \cite{langer_ambegaokar,zhang}. The boundary condition for the wire is $kL=2\pi n$, where $L$ is the wire length, and for a ring $kL+2\pi \phi=2\pi n$, where $L=2\pi R$. When biasing a wire with current $I<I_\mathrm{c}$, $k$ is not uniquely determined  since $I \propto k(1-k^2)$ has multiple solutions,  and the system will always chose the value of $k$ in the stable region (non-shaded area in Supplementary Figure \ref{free_energy}b-d). (Here "stable" refers to the long wire/ring diameter limit). In contrast, when biasing a ring with flux, $k$ is uniquely  determined (through the boundary condition), and therefore it is possible to bias the system in the shaded region, which corresponds to the regions indicated by black arrows in Figure 4 in the main text. In these regions the velocity is super-critical (see Supplementary Figure \ref{free_energy}d), but the diminishing density leads to the decrease of current. This effect is only accessible in the the ring configuration.

\subsection{Supplementary Note 7. Phase slip flux}

In Supplementary Figure \ref{switching_sq} we show the extended version of Figure 3 from the main text. Data for each ring size is given in a separate panel. The measured transition widths are denoted by bars. Dotted lines are the theoretically predicted values of the switching flux in the limit of a long sample (Eq. (1) in the main text) and full lines are the theory prediction which takes into account the finite-length correction (Eq. (2) in the main text).

\begin{figure}[h!]
\centerline{\hbox{
\epsfig{figure=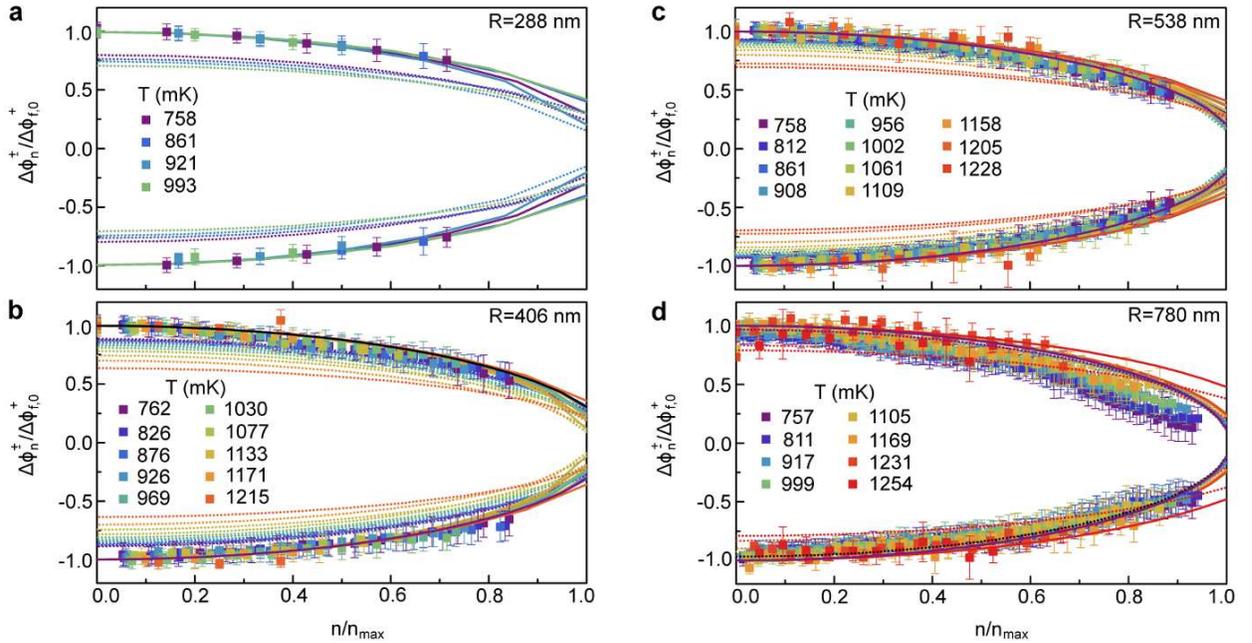,width=166mm}}} 
\caption{\textbf{Phase slip flux as function of winding number.} Dots: experimental values; bars: observed width of each jump; full lines: prediction for the switching flux $\Delta\phi_{\mathrm{f},n}^{\pm}$; dotted lines: prediction for the switching flux $\Delta\phi_{\mathrm{c},n}^{\pm}$ (see main text). Colors represent temperature. Radii $R=288,\:406,\:538,\:780$ nm of four different samples are denoted on the panels. The normalization of the axes is explained in the main text. \break}
\label{switching_sq}
\end{figure}

The largest disagreement between data and prediction occurs for the largest rings at low temperature ($R=780$ nm, blue and violet squares in Supplementary Figure \ref{switching_sq}d). This discrepancy is likely due to the increased importance of the rings' self-inductance in this regime, which is ignored in our analysis. Self-inductance leads to the skewing of the current-phase relation and as consequence the GL fit doesn't work as well.

\subsection{Supplementary Note 8. Transition width}

We have also studied the width of the jumps from one winding number to another, which is non-zero since the measurement is performed on an array of rings. The result is shown in Supplementary Figure \ref{width} where the transition widths are given as function of the winding number $n$, proportional to $B$, for all ring sizes and temperatures.
We see that the transition widths have a non-zero value at $n=0$ and increase roughly linearly with $B$. The slope of the $B$-dependence is independent of $T$ and decreases with $R$.  It is consistent with lithographic ring-to-ring imprecision $\Delta R=$ 1.9, 1.5, 2.1 and 2.0 nm for rings with $R=$ 780, 538, 406 and 288 nm respectively.

\begin{figure}[h]
\centerline{\hbox{
\epsfig{figure=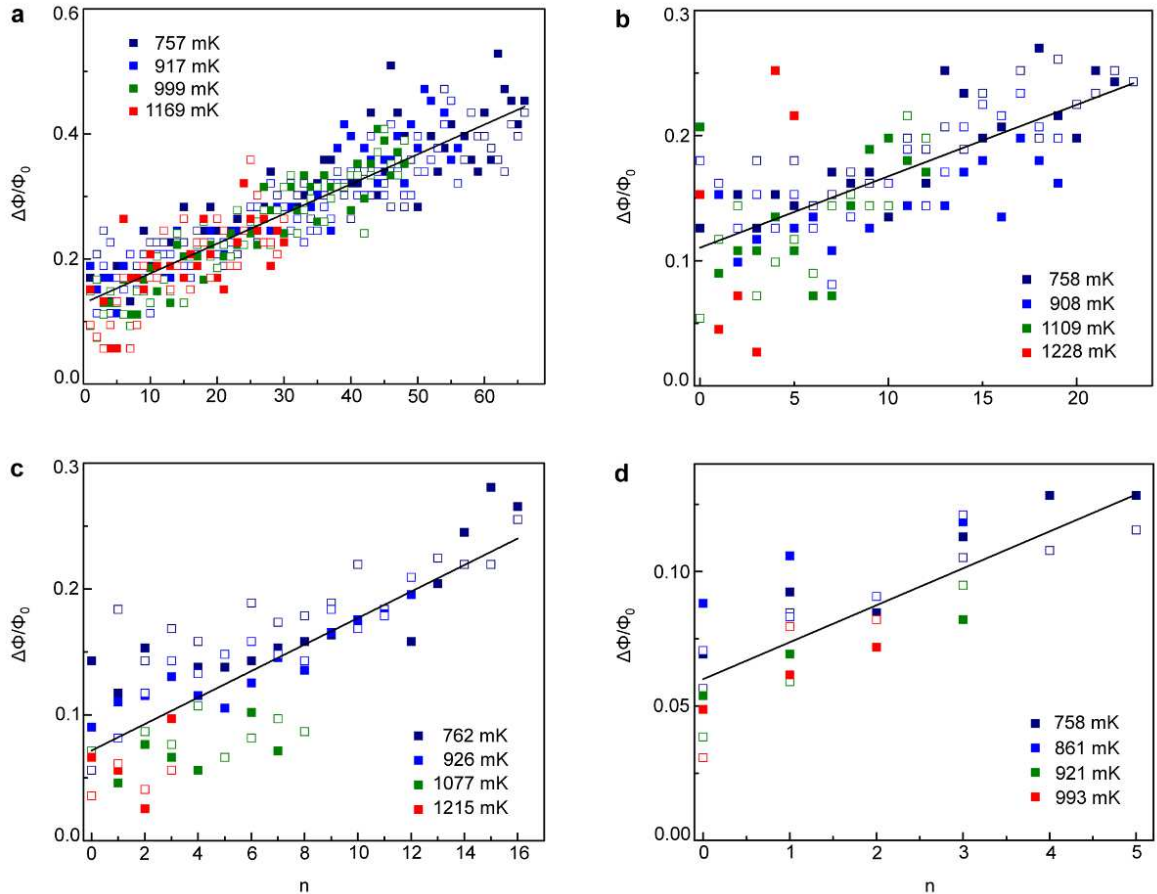,width=156mm}}
}
\caption{\textbf{Transition width.} Observed transition width in units of normalized flux $\Delta\Phi/\Phi_0$ as function of winding number $n$, for different ring sizes and temperatures. Solid lines indicate a linear fit to the whole data set at each ring size. Radius sizes are: (\textbf{a}) $R=780$ nm, (\textbf{b}) $R=538$ nm, (\textbf{c}) $R=406$ nm and (\textbf{d}) $R=288$ nm.
\break }
\label{width}
\end{figure}

The transition width at zero field also shows no discernible temperature dependence. It is a factor of $5-10$ larger than expected from the rings' mutual inductance, which results in rings at the middle of an array seeing a slightly different field than those at the edge.
Extrapolating the linear behavior of the transition width to zero field gives $\Phi_{n=0}$, which is seen to increase with $R$.
	
We conclude from these observations that the rings' temperature does not influence the transition widths of the arrays. This is consistent with the fact that the transition width expected for thermal switching across a barrier \cite{mccumber_halperin} is estimated to be several times less than the observed width.

\subsection{Supplementary Note 9. Estimate of damping at low temperatures}

As explained in the main text, the presence of a large supercurrent justifies  the use of the time-dependent Ginzburg-Landau theory at $T\to T_{\mathrm{c}}$. In that approximation, the time evolution of the phase difference across the phase slip is described by viscous "motion" of the  phase. As a result, $\Delta n=1$ in a deterministic phase slip at $T$ close to $T_{\mathrm{c}}$. Since there is no quantitative theory of deterministic phase slips at low temperature, we proceed with estimates helping to asses the possibility of $\Delta n>1$.

At high winding numbers, the dependence of the kinetic energy of the moving condensate on current $I$ can be approximated by the linear function  $\delta E=(\hbar/e) I$. The deterministic phase slip occurs once $I$ reaches its critical value, $I=I_\mathrm{c}$. Therefore, in a phase slip the condensate energy changes by $E_{\Delta n}=(\Delta n \hbar /e)I_\mathrm{c}$. The critical current $I_\mathrm{c}=S j_\mathrm{c}$ is proportional to the cross-sectional area $S$ of the wire making the ring. The critical current density $j_\mathrm{c}$ can be estimated from the Usadel~\cite{usadel} and BCS self-consistency equations~\cite{tinkham} that allow one to relate $j_\mathrm{c}$ to the gap width $\Delta$, the electron diffusion coefficient, and the electron density of states at the Fermi level. As a result, at low temperatures $j_\mathrm{c}\sim\Delta/(e\rho\,\xi)$ is expressed in terms of the corresponding values of $\Delta$ and coherence length $\xi$, and the normal-state resistivity $\rho$ (we note that $j_\mathrm{c}$ acquires an additional factor $(1-T/T_\mathrm{c})^{3/2}$ at $T\to T_\mathrm{c}$, in accordance with the GL theory). Thus, at low temperatures we find for the condensate energy difference
\begin{equation}
E_{\Delta n=1}\sim \frac{\hbar}{e^2}\frac{S}{\rho\,\xi}\Delta\,.
\label{E1}
\end{equation}
The second factor here, $G=S/(\rho\,\xi)$, has the meaning of the normal-state conductance of a wire segment long enough to house a phase slip.

The barrier "protecting" the lower metastable state is $\delta F = (\hbar/e)I_\mathrm{c} \sqrt{3/2} \left(1-j/j_\mathrm{c}\right)^{5/4} = E_{\Delta n = 1} \sqrt{3/2} \left(1-j/j_\mathrm{c}\right)^{5/4} $ \cite{langer_ambegaokar}. As already mentioned in Supplementary Note 6, $I \sim j \sim k(1-k^2)$, where $k$ is the  wave-vector of the superconducting phase (here in the units of $1/\xi$). Close to $I_\mathrm{c}$ ($k\rightarrow 1/\sqrt{3}$), we have $1-j/j_\mathrm{c} \sim (9/2)\left(k-1/\sqrt{3}\right)^2$, which, in combination with $k=(\xi/R)(n-\phi)$, yields $\delta F_{\Delta n=1} \sim \left(\xi/R\right)^{5/2} E_{\Delta n=1}$.

To estimate the dissipation due to the quasiparticle production in the course of a phase slip, we model it as a short SNS junction of conductance $G$. A finite phase difference $\varphi$ across it results in the appearance of Andreev levels with sub-gap energies. Furthermore, the time dependence of $\varphi$ leads to Landau-Zener transitions between the occupied and empty levels. As the result, an out-of-equilibrium level occupation is created. In a $\Delta n=1$ phase slip, the phase difference $\varphi$ starts from $0$ and ends at $2\pi$; respectively, the Andreev levels "peel off" and merge with the edges of the gap $\Delta$. At the end of the cycle, the non-equilibrium occupation of levels transforms into pairs of free quasiparticles each pair carrying energy $2\Delta$. The number of created pairs $N_p$ depends on $d\varphi/d\tau$ ($\tau$ denotes time) in the course of the phase slip. To estimate it, we use the result~\cite{averin} developed for the dissipative current across a short SNS junction, $I_{\rm diss}=G\sqrt{|V|\Delta/e}$ at a constant low ($eV\ll\Delta$) bias, $V=(\hbar/2e)(d\varphi/d\tau)$. For estimates, we set $|d\varphi/d\tau|\sim2\pi/\tau_\mathrm{ps}$ with $\tau_\mathrm{ps}$ being the time it takes to undergo a phase slip. Dispensing with the unreliable numerical factors, the energy spent on the quasiparticles' production can be estimated as

\vspace{-2mm}

\begin{equation}
E_{\rm diss}=2\Delta N_p\sim \Delta \tau_\mathrm{ps} I_{\rm diss}/e
\sim\frac{\hbar}{e^2}G\Delta\left(\Delta\tau_\mathrm{ps}/\hbar\right)^{1/2}\,.
\label{Ediss}
\end{equation}

\noindent Lastly, we use the estimate $\tau_\mathrm{ps}\sim\hbar/\Delta$. It may be viewed as the extrapolation of the TDGL characteristic time $\sim\hbar/|T-T_\mathrm{c}|$ to low temperatures, or as the $RC$ time constant of the junction with the capacitance $C\sim \hbar G/\Delta$ renormalized by the quantum fluctuations of charge~\cite{larkin-ovchinnikov}; the two approaches yield the same result. Replacing $\tau_\mathrm{ps}\to\hbar/\Delta$ and using the conductance $G=S/(\rho\,\xi)$ associated with the phase slip in Supplementary Equation ~(\ref{Ediss}), we find $E_{\rm diss}\sim (\hbar S/e^2\rho\,\xi)\Delta$ quoted in the main text.

\subsection{Supplementary References}


\begin{thebibliography}{99}


\bibitem{little} Little, W. A. Decay of persistent  currents in small superconductors. \emph{Phys. Rev.} \textbf{156}, 396-403 (1967).

\bibitem{langer_ambegaokar} Langer, J. S. \& Ambegaokar, V. Intrinsic resistive transition in narrow
superconducting channels.  \emph{Phys. Rev.} \textbf{164}, 498-510 (1967).

\bibitem{mccumber_halperin} McCumber, D. E. \& Halperin, B. I. Time scale of intrinsic resistive fluctuations in thin superconducting wires. \emph{Phys. Rev. B} \textbf{1}, 1054-1070 (1970).

\bibitem{tinkham} Tinkham, M. \emph{Introduction to Superconductivity} 2nd edn (Dover, 2004).

\bibitem{halperin} Halperin, B. I., Refael, G. \& Demler, E. Resistance in superconductors. \emph{Int. J. Mod. Phys. B} \textbf{24}, 4039-4080 (2010).

\bibitem{tarlie_elder} Tarlie, M. B. \& Elder, K. R. Metastable state selection in one-dimensional systems with a time-ramped control parameter. \emph{Phys. Rev. Lett.} \textbf{81}, 18-21 (1998).


\bibitem{giordano} Giordano, N. Evidence for macroscopic quantum tunneling in one-dimensional superconductors. \emph{Phys. Rev. Lett. } \textbf{61}, 2137-2140 (1988).

\bibitem{duan} Duan, J. M. Quantum decay of one-dimensional supercurrent: role of electromagnetic field. \emph{Phys. Rev. Lett.} \textbf{74}, 5128-5131 (1995).

\bibitem{zaikin} Zaikin, A. D., Golubev, D. S., van Otterlo, A. \& Zim\'{a}nyi, G. T. Quantum phase slips and transport in ultrathin superconducting wires. \emph{Phys. Rev. Lett.} \textbf{78}, 1552-1555 (1997).

\bibitem{golubov_zaikin} Golubev, D. S. \& Zaikin, A. D. Quantum tunneling of the order parameter in superconducting nanowires. \emph{Phys. Rev. B} \textbf{64}, 014504 (2001).

 \bibitem{bradley_doniach} Bradley, R. M. \& Doniach, S. Quantum fluctuations in chains of Josephson junctions. \emph{Phys. Rev B} \textbf{30}, 1138-1147 (1984).

\bibitem{fazio_vanderzant} Fazio, R. \& van der Zant, H. Quantum phase transitions and vortex dynamics in superconducting networks. \emph{Physics Reports} \textbf{355}, 235-334 (2001).

\bibitem{matveev} Matveev, K. A., Larkin, A. I. \& Glazman, L. I. Persistent current in superconducting nanorings. \emph{Phys. Rev. Lett.} \textbf{89}, 096802 (2002).

\bibitem{buchler} B\"{u}chler, H. P., Geshkenbein, V. B. \& Blatter, G. Quantum fluctuations in thin superconducting wires of finite length. \emph{Phys. Rev. Lett.} \textbf{92}, 067007 (2004).

\bibitem{taps_exp} Newbower, R. S., Beasley, M. R. \& Tinkham, M. Fluctuation effects on the superconducting transition of tin whisker crystals. \emph{Phys. Rev. B} \textbf{5}, 864-868 (1972).

\bibitem{qps_exp1} Giordano, N. \& Schuler, E. R. Macroscopic quantum tunneling and related effects in a one-dimensional superconductor. \emph{Phys. Rev. Lett.} \textbf{63}, 2417-2420 (1989).

\bibitem{qps_exp2}  Bezryadin, A., Lau, C. N. \&  Tinkham, M. Quantum suppression of superconductivity in ultrathin nanowires. \emph{Nature} \textbf{404}, 971-974 (2000).

\bibitem{qps_exp3}  Lau, C. N., Markovic, N., Bockrath, M., Bezryadin, A. \& Tinkham, M. Quantum phase slips in superconducting nanowires. \emph{Phys. Rev. Lett.} \textbf{87}, 217003 (2001).

\bibitem{qps_exp7}  Altomare, F., Chang, A. M., Melloch, M. R., Hong, Y. \& Tu, C. W. Evidence for macroscopic quantum tunneling of phase slips in long one-dimensional superconducting Al wires. \emph{Phys. Rev. Lett.} \textbf{97}, 017001 (2006).

\bibitem{qps_exp8}  Sahu, M. \emph{et al}. Individual topological tunnelling events of a quantum field probed through their macroscopic consequences. \emph{Nat. Phys.} \textbf{5}, 503-508 (2009).

\bibitem{qps_exp9}  Li, P., Wu, P. M., Bomze, Y., Borzenets, I. V., Finkelstein, G. \& Chang, A. M. Switching currents limited by single phase slips in one-dimensional superconducting Al nanowires. \emph{Phys. Rev. Lett.} \textbf{107}, 137004 (2011).

\bibitem{qps_exp10} Aref, T., Levchenko, A., Vakaryuk, V. \&  Bezryadin, A. Quantitative analysis of quantum phase slips in superconducting Mo$_76$Ge$_24$ nanowires revealed by switching-current statistics. \emph{Phys. Rev. B} \textbf{86}, 024507 (2012).

\bibitem{artyunov} Arutyunov, K. Yu., Hongisto, T. T., Lehtinen, J. S., Leino, L. I., \& Vasiliev, A. L. Quantum phase slip phenomenon in ultra-narrow superconducting nanorings. \emph{Sci. Rep.} \textbf{2}, 293(1-7) (2012).

\bibitem{belkin_tech} Belkin, A., Brenner, M., Aref, T., Ku, J., \& Bezryadin, A. Little-Parks oscillations at low temperatures: Gigahertz resonator method. \emph{Appl. Phys. Lett.} \textbf{98}, 242504 (2011).

\bibitem{belkin} Belkin, A., Belkin, M., Vakaryuk, V., Khlebnikov, S. \& Bezryadin, A. Formation of quantum phase slip pairs in superconducting nanowires. \emph{Phys. Rev. X} \textbf{5}, 021023 (2015).

\bibitem{zhang} Zhang X. \& Price, J. C. Susceptibility of a mesoscopic superconducting ring. \emph{Phys. Rev. B} \textbf{55}, 3128-3140 (1997).

\bibitem{bourgeois} Bourgeois, O., Skipetrov, S. E., Ong, F. \& Chaussy, J. Attojoule calorimetry of mesoscopic superconducting loops. \emph{Phys. Rev. Lett.} \textbf{94}, 057007 (2005).

\bibitem{bert_moler} Bert, J. A., Koshnick, N. C., Bluhm, H. \& Moler, K. A. Fluxoid fluctuations in mesoscopic superconducting rings. \emph{Phys. Rev. B} \textbf{84}, 134523 (2011).

\bibitem{koshnick_moler} Koshnick, N. C., Bluhm, H., Huber, M. E. \& Moler, K. A. Fluctuation superconductivity in mesoscopic aluminum rings. \emph{Science} \textbf{318}, 1440-1443 (2007).

\bibitem{pedersen} Pedersen,  S., Kofod, G. R., Hollingbery, J. C., S{\o}rensen, C. B. \& Lindelof, P. E. Dilation of the giant vortex state in a mesoscopic superconducting loop. \emph{Phys. Rev. B} \textbf{64}, 104522 (2001).

\bibitem{vodolazov} Vodolazov, D. Y., Peeters, F. M., Dubonos, S. V. \& Geim, A. K. Multiple flux jumps and irreversible behavior of thin Al superconducting rings. \emph{Phys. Rev. B} \textbf{67}, 054506 (2003).

\bibitem{bluhm_moler} Bluhm, H., Koshnick, N. C., Huber, M. E. \& Moler, K. A. Magnetic response of mesoscopic superconducting rings with two order parameters. \emph{Phys. Rev. Lett.} \textbf{97}, 237002 (2006).

\bibitem{tedrow_meservey} Tedrow, P. M. \& Meservey, R. Spin-paramagnetic effects in superconducting aluminum films. \emph{Phys. Rev. B} \textbf{8}, 5098-5108 (1973).

\bibitem{kramer_zimmermann} Kramer L. \& Zimmermann, W. On the Eckhaus instability for spatially periodic patterns. \emph{Physica D} \textbf{16}, 221-232 (1985).

\bibitem{tuckerman} Tuckerman L. S. \& Barkley, D. Bifurcation analysis of the Eckhaus instability. \emph{Physica  } \textbf{46D}, 57-86 (1990).

\bibitem{ania_science} Bleszynski-Jayich, A. C. \emph{et al.} Persistent currents in normal metal rings. \emph{Science} \textbf{326}, 272-275 (2009).

\bibitem{will_thesis} Shanks, W. E. Persistent currents in normal metal rings. \emph{Thesis} (Yale University, 2011).

\bibitem{manuel} Castellanos-Beltran, M. A., Ngo, D. Q., Shanks, W. E., Jayich, A. B. \& Harris, J. G. E. Measurement of the full distribution of persistent current in normal-metal rings. \emph{Phys. Rev. Lett.} \textbf{110}, 156801 (2013).

\bibitem{little_parks} Little, W. A. \& Parks, R. D.  Observation of quantum periodicity in the transition temperature of a superconducting cylinder. \emph{Phys. Rev. Lett.} \textbf{9}, 9-12 (1962).

\bibitem{little_parks2}  Parks, R. D. \& Little, W. A. Fluxoid quantization in a multiply-connected superconductor. \emph{Phys. Rev.} \textbf{133}, A97-A103 (1964).

\bibitem{pearl}  Pearl, J. Current distribution in superconducting films carrying quantized fluxoids. \emph{Appl. Phys. Lett.} \textbf{5}, 65-66 (1964).


\bibitem{maloney} Maloney, M. D., de la Cruz, F. \& Cardona, M. Superconducting parameters and size effects of aluminum films and foils. \emph{Phys. Rev. B} \textbf{5}, 3558-3572 (1972).


\bibitem{levchenko-kamenev} Levchenko, A. \& Kamenev, A. Keldysh Ginzburg-Landau action of fluctuating superconductors. \emph{Phys. Rev. B} \textbf{76}, 094518 (2007).

\bibitem{averin} Bardas, A. \& Averin, D. V. Electron transport in mesoscopic disordered superconductor -- normal - metal -- superconductor junctions. \emph{Phys. Rev. B} \textbf{56}, R8518-R8521 (1997).


\bibitem{astafiev1} Astafiev, O.V. \emph{et al}. Coherent quantum phase slip. \emph{Nature} \textbf{484}, 355-358 (2012).

\bibitem{astafiev2} Peltonen, J. T. \emph{et al}. Coherent flux tunneling through NbN nanowires. \emph{Phys. Rev. B} \textbf{88}, 220506(R) (2013).

\bibitem{mooij_nazarov} Mooij, J. E. \& Nazarov, Yu. V. Superconducting nanowires as quantum phase-slip junctions. \emph{Nat. Phys.} \textbf{2}, 169-172 (2006).

\bibitem{mooij_harmans} Mooij, J. E. \& Harmans, C. J. P. M. Phase-slip flux qubits. \emph{New J. Phys.}  \textbf{7}, 219 (2005).


\end{thebibliography}

\begin{thebibliography}{99}

\bibitem{ania_science} Bleszynski-Jayich, A. C. \emph{et al.} Persistent currents in normal metal rings. \emph{Science} \textbf{326}, 272-275 (2009).

\bibitem{wills_thesis} Shanks, W. E. Persistent currents in normal metal rings. \emph{Thesis} (Yale University, 2011).

\bibitem{little_parks} Little, W. A. \& Parks, R. D.  Observation of quantum periodicity in the transition temperature of a superconducting cylinder. \emph{Phys. Rev. Lett.} \textbf{9}, 9-12 (1962).

\bibitem{little_parks2}  Parks, R. D. \& Little, W. A. Fluxoid quantization in a multiply-connected superconductor. \emph{Phys. Rev.} \textbf{133}, A97-A103 (1964).

\bibitem{tinkham} Tinkham, M. \emph{Introduction to Superconductivity} 2nd edn (Dover, 2004).

\bibitem{zhang} Zhang X. \& Price, J. C. Susceptibility of a mesoscopic superconducting ring. \emph{Phys. Rev. B} \textbf{55}, 3128-3140 (1997).

\bibitem{fink_grunfeld} Fink, H. J. \& Gr\"{u}nfeld, V. Flux periodicity in superconducting rings: Comparison to loops with Josephson junctions. \emph{Phys. Rev. B} \textbf{33}, 6088-6093 (1986).

\bibitem{langer_ambegaokar} Langer, J. S. \& Ambegaokar, V. Intrinsic resistive transition in narrow
superconducting channels.  \emph{Phys. Rev.} \textbf{164}, 498-510 (1967).

\bibitem{mccumber_halperin} McCumber, D. E. \& Halperin, B. I. Time scale of intrinsic resistive fluctuations in thin superconducting wires. \emph{Phys. Rev. B} \textbf{1}, 1054-1070 (1970).

\bibitem{usadel}Usadel, K. D. Generalized Diffusion Equation for Superconducting Alloys. \emph{Phys. Rev. Lett.} \textbf{25}, 507-509 (1970).

\bibitem{averin} Bardas, A. \& Averin, D. V. Electron transport in mesoscopic disordered superconductor -- normal - metal -- superconductor junctions. \emph{Phys. Rev. B} \textbf{56}, R8518-R8521 (1997).

\bibitem{larkin-ovchinnikov} Larkin, A. I. \& Ovchinnikov, Yu. N. Decay of the supercurrent in tunnel junctions. \emph{Phys. Rev. B} \textbf{28}, 6281-6285 (1983).

\end{thebibliography}
\end{document}